\documentclass[aps,prd,twocolumn,groupedaddress,nofootinbib,amssymb,eqsecnum,showpacs,epsfig]{revtex4}
\usepackage{graphicx}
\usepackage{bm}
\usepackage{dcolumn}
\usepackage{amsmath}
\usepackage{hyperref}

\def \lleq {\lower0.9ex\hbox{ $\buildrel < \over \sim$} ~}
\def \ggeq {\lower0.9ex\hbox{ $\buildrel > \over \sim$} ~}

\def \om   {\Omega_{\rm 0m}}

\def \beq  {\begin{equation}}
\def \eeq  {\end{equation}}
\def \ber  {\begin{eqnarray}}
\def \eer  {\end{eqnarray}}

\begin{document}
\newcommand{\newc}{\newcommand}

\newcommand{\ben}{\begin{eqnarray}}
\newcommand{\een}{\end{eqnarray}}
\newc{\be}{\begin{equation}}
\newc{\ee}{\end{equation}}
\newc{\ba}{\begin{eqnarray}}
\newc{\ea}{\end{eqnarray}}
\newc{\bea}{\begin{eqnarray*}}
\newc{\eea}{\end{eqnarray*}}
\newc{\D}{\partial}
\newc{\ie}{{\it i.e.} }
\newc{\eg}{{\it e.g.} }
\newc{\etc}{{\it etc.} }
\newc{\etal}{{\it et al.}}
\newcommand{\nn}{\nonumber}
\newc{\ra}{\rightarrow}
\newc{\lra}{\leftrightarrow}
\newc{\lsim}{\buildrel{<}\over{\sim}}
\newc{\gsim}{\buildrel{>}\over{\sim}}
\title{Genetic algorithms and supernovae type Ia analysis}
\author{Charalampos Bogdanos$^1$}
\email{Charalampos.Bogdanos@th.u-psud.fr}
\author{Savvas Nesseris$^2$}
\email{nesseris@nbi.dk}

\affiliation{$^1$ LPT, Universit$\acute{e}$ de Paris-Sud-11,
B$\hat{a}$t. 210, 91405 Orsay CEDEX, France \\ $^2$ The Niels Bohr
International Academy, The Niels Bohr Institute, Blegdamsvej-17,
DK-2100, Copenhagen \O, Denmark}
\date{\today}

\begin{abstract}
We introduce genetic algorithms as a means to analyze supernovae
type Ia data and extract model-independent constraints on the
evolution of the Dark Energy equation of state $w(z)\equiv
\frac{P_{DE}}{\rho_{DE}}$. Specifically, we will give a brief
introduction to the genetic algorithms along with some simple
examples to illustrate their advantages and finally we will apply
them to the supernovae type Ia data. We find that genetic
algorithms can lead to results in line with already established
parametric and non-parametric reconstruction methods and could be
used as a complementary way of treating SNIa data. As a
non-parametric method, genetic algorithms provide a
model-independent way to analyze data and can minimize bias due to
premature choice of a dark energy model.
\end{abstract}

\pacs{}

\maketitle

\section{Introduction}
The accelerated expansion of the universe has been confirmed
during the last decade by several observational probes
\cite{Riess:2004nr,Spergel:2006hy,Readhead:2004gy,Goldstein:2002gf,Rebolo:2004vp,Tegmark:2003ud,Hawkins:2002sg}.
The origin of this acceleration may be attributed to either dark
energy with negative pressure, or to a modification of General
Relativity that makes gravity repulsive at recent times on
cosmological scales. One way to distinguish between these two
possibilities and identify in detail the gravitational properties
of dark energy or modified gravity is to have a detailed mapping
of the expansion rate $H(z)$ as a function of the redshift $z$.
This is equivalent to identifying the dark energy equation of
state $w(z)\equiv \frac{P}{\rho}$ which can be written as \be
w(z)\,=\,-1\,+\frac{1}{3}(1+z)\frac{d\ln (\delta H(z)^2)}{d\ln
z}\,,{\label{wzh1}} \ee where $\delta
H(z)^2=H(z)^2/H_0^2-\Omega_{\rm 0m} (1+z)^3$ accounts for all
terms in the Friedmann equation not related to matter. The
cosmological constant ($w(z)=-1$) corresponds to a constant dark
energy density, while in general it can be time dependent.

It has been shown \cite{Vikman:2004dc} that a $w(z)$ observed to
cross the line $w(z)=-1$ (phantom divide line) is very hard to
accommodate in a consistent theory in the context of General
Relativity. On the other hand, such a crossing can be easily
accommodated in the context of extensions of General Relativity
\cite{Boisseau:2000pr}. Therefore, the crossing of the phantom
divide line $w=-1$ could be interpreted as a hint in the direction
of modified gravity, see for example \cite{Boisseau:2000pr},
\cite{Bamba:2008hq},\cite{Nesseris:2006er}. Such a hint would
clearly need to be verified by observations of linear density
perturbation evolution through e.g. weak lensing
\cite{Refregier:2006vt} or the redshift distortion factor
\cite{Hamilton:1997zq}.

Early SNIa data put together with more recent such data through
the Gold dataset \cite{Riess:2004nr,Riess:2006fw} have been used
to reconstruct $w(z)$, and have demonstrated a mild preference for
a $w(z)$ that crossed the phantom divide line
\cite{Nesseris:2005ur,Nesseris:2006er}. A cosmological constant
remained consistent but only at the $2\sigma$ level. However, the
Gold dataset has been shown to suffer from systematics due to the
inhomogeneous origin of the data \cite{Nesseris:2006ey}. More
recent SNIa data (SNLS\cite{Astier:2005qq},
ESSENCE\cite{WoodVasey:2007jb}, HST\cite{Riess:2006fw})
re-compiled in \cite{Davis:2007na} have demonstrated a higher
level of consistency with $\Lambda$CDM and showed no trend for a
redshift dependent equation of state.

However, despite the recent progress the true nature of Dark
Energy still remains a mystery with many possible candidates being
investigated, see for example \cite{Perivolaropoulos:2006ce}. The
simplest possible candidate assumes the existence of a positive
cosmological constant which is small enough to have started
dominating the universe at recent times. This model provides an
excellent fit to the cosmological observational data
\cite{Komatsu:2008hk} and has the additional bonus of simplicity
and a single free parameter. Despite its simplicity and good fit
to the data, this model fails to explain why the cosmological
constant is so unnaturally small as to come to dominate the
universe at recent cosmological times, a problem known as the {\it
coincidence problem} and there are specific cosmological
observations which differ from its predictions
\cite{Perivolaropoulos:2008ud},\cite{Perivolaropoulos:2008yc}.

A further complication in the investigation of the behavior of
dark energy occurs due to the bias introduced by the
parameterizations used. At the moment, there is a multitude of
available phenomenological ans\"atze for the dark energy equation
of state parameter $w$, each with its own merits and limitations
(see \cite{Sahni:2006pa} and references therein). The
interpretation of the SNIa data has been shown to depend greatly
on the type of parametrization used to perform a data fit
\cite{Sahni:2006pa}. Choosing a priori a model for dark energy can
thus adversely affect the validity of the fitting method and lead
to compromised or misleading results. The need to counteract this
problem paved the way for the consideration of a complementary set
of non-parametric reconstruction techniques
\cite{Daly:2003iy,Wang:2003gz,Saini:2003pa}. These try to minimize
the ambiguity due to a possibly biased assumption for $w$ by
fitting the original datasets without using any parameters related
to some specific model. The result of these methods can then be
interpreted in the context of a DE model of choice. Non-parametric
reconstructions can thus corroborate parametric methods and
provide more credibility. However, they too suffer from a
different set of problems, mainly the need to resort to
differentiation of noisy data, which can itself introduce great
errors.

In this paper, we consider a method for non-parametric
reconstruction of the dark energy equation of state parameter $w$,
based on the notions of genetic algorithms and grammatical
evolution. We show how the machinery of genetic algorithms can be
used to analyze SNIa data and provide examples of its application
to determine dark energy parameters for specific models. Our
results are consistent with already known findings of both
parametric and non-parametric treatments. In particular, we obtain
parameters which agree within $1\sigma$ with $\Lambda CDM$, but
show a slight trend towards a varying parameter $w(z)$. As a
non-parametric method, genetic algorithms provide a
model-independent way to analyze data and can minimize bias due to
premature choice of a dark energy model.

A rather formal definition of the Genetic Algorithms (GA)
\cite{Goldberg} is that of a computer simulation in which a
solution to some predefined problem is searched for in a
stochastic way by evolving an ensemble (population) of candidate
solutions (usually called chromosomes) towards an optimal state,
where some of the chromosomes have reached sufficiently close to
the true solution. This approach to problem solving resembles the
evolutionary pattern encountered in nature, where an optimal or
dominant organism emerges through a series of apparently random
mutations and combinations between different individuals. The
concept of optimality of a chromosome is usually encoded in the
genetic algorithm language in a {\it fitness function},
corresponding to the phenotype of an individual. Candidate
solutions with a higher fitness function are considered to be
better than the rest of the population, in the sense that they are
closer to the true solution. Genetic operators such as mutation
and combination (crossover) are then implemented to produce
offspring from a given population with better properties.
Chromosomes with a good fitness are usually more likely to
crossover and produce descendants and the combination often takes
place between the best candidate solutions of a generation. It is
thus more likely for descendants to bear a combination of already
favorable genes, leading to an improvement in fitness. The genetic
algorithm method is in a way similar to search algorithms, such as
hill climbing, $A^{*}$\cite{alphastar} or simulated
annealing\cite{annealing}.

GAs are more useful and efficient than usual techniques when
\begin{itemize}
    \item The parameter space is very large, too complex or not enough understood.
    \item Domain knowledge is scarce or expert knowledge is difficult to encode to narrow the search space.
    \item Traditional search methods give poor results or completely fail.
\end{itemize} so naturally they have been used with success in many fields where
one of the above situations is encountered, like the computational
science, engineering and economics. Recently, they have also been
applied to study high energy physics
\cite{Becks:1994mm},\cite{Allanach:2004my}, \cite{Rojo:2004iq}
gravitational wave detection \cite{Crowder:2006wh} and
gravitational lensing \cite{Brewer:2005ww}. Since the nature of
Dark Energy still remains a mystery, this makes it for us an ideal
candidate to use the GAs as a means to analyze the SNIa data and
extract model independent constraints on the evolution of the Dark
Energy equation of state $w(z)$. In Section 2 we will briefly
describe the standard way of analyzing SNIa data while a more
thorough treatment of GAs and grammatical evolution (GE) is
deferred to Section 3. Section 4 contains the general methodology
of the GA paradigm and its application to Supernovae data.

\section{The Standard $\textrm{SNIa}$ data analysis}

We use the SNIa dataset of Kowalski et. al. \cite{Kowalski:2008ez}
consisting of 414 SNIa, reduced to 307 points after various
selection cuts were applied in order to create a homogeneous and
high-signal-to-noise dataset. These observations provide the
apparent magnitude $m(z)$ of the supernovae at peak brightness
after implementing the correction for galactic extinction, the
K-correction and the light curve width-luminosity correction. The
resulting apparent magnitude $m(z)$ is related to the luminosity
distance $D_L(z)$ through \be m_{th}(z)={\bar M} (M,H_0) + 5
log_{10} (D_L (z))\,, \label{mdl} \ee where in a flat cosmological
model \be D_L (z)= (1+z) \int_0^z
dz'\frac{H_0}{H(z';\om,w_0,w_1)}\,, \label{dlth1} \ee is the
Hubble free luminosity distance ($H_0 d_L$),  and ${\bar M}$ is
the magnitude zero point offset and depends on the absolute
magnitude $M$ and on the present Hubble parameter $H_0$ as \ba
{\bar M} &=& M + 5 log_{10}\left(\frac{H_0^{-1}}{Mpc}\right) + 25= \nn \\
&=& M-5log_{10}h+42.38 \label{barm}. \ea The parameter $M$ is the
absolute magnitude which is assumed to be constant after the above
mentioned corrections have been implemented in $m(z)$.

The SNIa datapoints are given, after the corrections have been
implemented, in terms of the distance modulus \be
\mu_{obs}(z_i)\equiv m_{obs}(z_i) - M \label{mug}.\ee The
theoretical model parameters are determined by minimizing the
quantity \be \chi^2_{SNIa} (\om,w_0,w_1)= \sum_{i=1}^N
\frac{(\mu_{obs}(z_i) - \mu_{th}(z_i))^2}{\sigma_{\mu \; i}^2 }
\label{chi2}\,, \ee where $N=307$ and $\sigma_{\mu \; i}^2$ are
the errors due to flux uncertainties, intrinsic dispersion of SNIa
absolute magnitude and peculiar velocity dispersion. These errors
are assumed to be Gaussian and uncorrelated. The theoretical
distance modulus is defined as \be \mu_{th}(z_i)\equiv m_{th}(z_i)
- M =5 log_{10} (D_L (z)) +\mu_0\,, \label{mth} \ee where \be
\mu_0= 42.38 - 5 log_{10}h\,, \label{mu0}\ee and $\mu_{obs}$ is
given by (\ref{mug}). The steps we followed for the usual
minimization of (\ref{chi2}) in terms of its parameters are
described in detail in Refs.
\cite{Nesseris:2004wj,Nesseris:2005ur,Nesseris:2006er}.

This approach assumes that there is some theoretical model
available, given in the form of $H(z;p_i)$ and in terms of some
parameters $p_i$ like the $\om$ or the $w_0$ and $w_1$ of the CPL
ans\"atz, which is to be compared against the data. As a result of
the analysis, the best-fit parameter values and the corresponding
$1\sigma$ error bars are obtained. The problem with this method is
that the obtained values of the parameters are model-dependent and
in general models with more parameters tend to give better fits to
the data. This is where the GA approach starts to depart from the
ordinary parametric method. Our goal is to minimize a function
like (\ref{chi2}), not using a candidate model function for the
distance modulus and varying parameters, but through a stochastic
process based on a GA evolution. This way, no prior knowledge of a
dark energy models is needed to obtain a solution. The resulting
``theoretical'' expression for $\mu_{th}(z)$ is completely
parameter-free. This is the main reason for the use of GA in this
paper. We now proceed to give a more thorough treatment of GAs and
the results of their application in the next sections.

\section{Genetic algorithms}
\subsection{Overview}
GAs were introduced as a computational analogy of adaptive
systems. They are modelled loosely on the principles of the
evolution via natural selection, employing a population of
individuals that undergo selection in the presence of
variation-inducing operators such as  mutation and crossover. The
encoding of the chromosomes is called the genome or genotype and
is composed, in loose correspondence to actual DNA genomes, by a
series of representative ``genes''. Depending on the problem, the
genome can be a series (vector) of binary numbers, decimal
integers, machine precision integers or reals or more complex data
structures. As mentioned in the introduction, a fitness function
is used to evaluate individual chromosomes and reproductive
success usually varies with fitness. The fitness function is a map
between the gene sequence of the chromosomes (genotype) and a
number of attributes (phenotype), directly related to the
properties of a wanted solution. Often the fitness function is
used to determine a ``distance'' of a candidate solution from the
true one. Various distance measures can be used for this purpose
(Euclidean distance, Manhattan, Mahalanobis etc.).

The algorithm begins with an initial population of candidate
solutions, which is usually randomly generated. Although GA's are
relatively insensitive to initial conditions, i.e. the population
we start from is not very significant, using some prescription for
producing this seed generation can affect the speed of
convergence. In each successive step, the fitness functions for
the chromosomes of the population are evaluated and a number of
genetic operators (mutation and crossover) are applied to produce
the next generation. This process continues until some termination
criteria is reached, e.g. obtain a solution with fitness greater
than some predefined threshold or reach a maximum number of
generations. The later is imposed as a condition to ensure that
the algorithm terminates even if we cannot get the desired level
of fitness.

The various steps of the algorithm can be summarized as follows:
\begin{enumerate}
    \item Randomly generate an initial population $M(0)$
    \item Compute and save the fitness for each individual m in the current population
    $M(t)$.
    \item Define selection probabilities $p(m)$ for each individual $m$ in $M(t)$ so that $p(m)$ is proportional to
    the fitness.
    \item Generate $M(t+1)$ by probabilistically selecting individuals
    from $M(t)$ to produce offspring via genetic
    operators.
    \item Repeat step 2 until satisfying solution is obtained, or maximum number of generations reached.
\end{enumerate}
The paradigm of GAs descibed above is usually the one applied to
solving most of the problems presented to GAs. Though it might not
find the best solution, more often than not, it would come up with
a partially optimal solution.

\subsection{Grammatical evolution}
The method of grammatical evolution (GE) was first introduced as
an alternative approach to genetic programming\cite{ONeil,ONeil2}.
Genetic programming is a variation of GAs, which uses programming
trees as chromosomes instead of vectors of binary numbers or
strings. These chromosomes can be used to express formal
expressions, mathematical functions or even entire programs
written in some specific programming language. GA operators such
as mutation or crossover can then be interpreted as operations
which alter or exchange subtrees of a chromosome tree, or splice
together two distinct trees.

Grammatical evolution can also be used to produce arbitrary
programs in any programming language but without the need for tree
structures, which usually require complicated and cumbersome
methods to implement genetic operators. Instead, one can use GE to
produce an expression or program using an ordinary binary vector
chromosome and a generative grammar, which maps the chromosome to
an expression using a set of grammatic rules. These rules are
usually expressed in a Backus-Naur form (BNF), which can be used
to recursively convert a binary vector into a string of terminal
symbols, which corresponds to the output. In this way, GE
introduces an additional layer of abstraction between the genome
(vector of binaries) and the phenotype (fitness). Usually, the
transition between the two is accomplished in one step, by
evaluating the fitness function of a chromosome. In the case of
grammatical evolution, there is the additional step of converting
the initial chromosome into an expression using the available
grammar. The expression is then evaluated to find the fitness of
the chromosome. GA operators can be easily implemented in this
scheme on the level of vector chromosomes, since the formal
expressions are evaluated in the next step and are not necessary
to perform the mutation and crossover. The added abstraction layer
also is more similar to what occurs in nature, where a
transcription process is used to map DNA base sequences (genotype)
into proteins (phenotype). GE has already been applied to problem
such as symbolic regression\cite{ONeil2}, minimization\cite{Tsoulos}, robot control\cite{Collins} and
finance\cite{Brabazon}.

\subsection{Initialization}
The first stage of the algorithm involves the creation of an
initial population. The chromosomes are generated in a random way
and in large enough quantities. The size of the population is very
important to successfully mimic a natural selection process, as
diversity is crucial to produce a variety of genes and phenotypes.
It is thus advisable to use as large a population sample as
possible in a simulation, while keeping computational complexity
under control. Usually the chromosome count remains fixed
throughout the evolutionary phase. Typically this involves a few
hundreds, up to several thousands of individuals, although these
numbers vary with the type of the problem and the resources
available. Similarly, the size of each chromosome (number of
genes) is going to affect the speed of the algorithm and the
quality of the final solution. Large genomes lead to richer
phenotypes and better chances to increase the fitness through
mutation and crossover, at the expense of computational
efficiency.

\subsection{Selection}
The purpose of this phase is to select the chromosomes in the
population which are going to breed to give a new generation. This
is the step where the equivalent of natural selection takes place.
Chromosomes are chosen for crossover based on their fitness.
Solutions with a better fitness function are the dominant members
of the population, hence they are more likely to be selected to
produce offspring.

Selection is performed in a stochastic way to avoid premature
convergence on suboptimal solutions. The algorithm assigns some
selection probability to each chromosome according to its fitness
function. The totality of the population or some part of it can be
used to draw a number of candidates for crossover. The most
popular and heavily used methods for selection are:
\begin{itemize}
\item Roulette wheel selection.
\item Tournament selection.
\end{itemize}
In the case of roulette wheel selection, a probability for
crossover of a chromosome is considered to be proportional to its
fitness. This is a widely used method, but suffers from a number
of drawbacks. Since chromosomes with higher fitness are more
likely to be selected for crossover, once a suboptimal solution
occurs, it may dominate the mating pool. At the same time, less
favored chromosomes tend to be completely neglected as candidates
for breeding. This leads to premature convergence to suboptimal
solutions and stagnation of the evolutionary computation. Roulette
wheel selection is also not so easy to implement.

On the other hand, tournament selection chooses the candidates for
crossover by taking smaller groups of randomly selected
individuals and picking the dominant member of each group. This is
an easy algorithm to implement and also more in line with what we
usually encounter in nature. Suboptimal chromosomes are thus more
likely to be picked than in the roulette wheel case and premature
convergence can be avoided more effectively. Tournament selection
is going to be used in our treatment of SNIa data.

\subsection{Reproduction}
After selecting the candidates for breeding, the next step is to
combine the corresponding genomes to produce the members of the
next generation. This is achieved by the application of genetic
operators, mutation and crossover.

For each pair of parent chromosomes picked from the population,
two children chromosomes are produced by exchanging the genomes of
the two parents. The easiest and more popular method is one-point
crossover, where a particular point is chosen on the genome
sequence and the genes from that point on completely exchanged
between the chromosomes. The descendants are then subjected to
mutation with some predefined probability (mutation rate). This
process of selection/crossover/mutation is then repeated until
enough individuals are produced to form a new generation. More
often the entire population of the previous generation is replaced
by members produced during the reproduction stage. In some cases,
elitism is applied to ensure that some of the already found
solution with the best fitnesses are passed on to the next
generation. Although the two parent scheme is more biology
inspired, recent researches suggested more than two parents may
lead to better quality chromosomes.

By crossover and mutation, a new generation with different members
from the progenitor is created. New chromosomes tend to inherent
the best qualities of the previous generation, since members with
higher fitness are more likely to be selected for reproduction. In
general, the new chromosomes will have better fitnesses on the
average and a better maximum fitness within the population.

\subsection{Termination}
The GA evolution continues until a termination
condition has been reached. Some of the most common conditions for termination include:

\begin{itemize}
    \item An optimal solution is found based on some best fitness threshold.
    \item The maximum number of generations is reached.
    \item Allocated budget (CPU time/money) reached.
    \item Further execution doesn't significantly improve the best fitness in the population.
    \item Combinations of the above.
\end{itemize}

\section{Results}
\subsection{General Methodology}
We first outline the course of action we follow to apply the GA
paradigm in the case of SNIa data. For the application of GA and
GE on the dataset, we use a modified version of the
GDF\cite{Tsoulos} tool
~\footnote{\href{http://cpc.cs.qub.ac.uk/summaries/ADXC}{http://cpc.cs.qub.ac.uk/summaries/ADXC}},
which uses GE as a method to fit datasets of arbitrary size and
dimensionality. This program uses the tournament selection method
for crossover. GDF requires a set of input data (train set), an
(optional) test sample and a grammar to be used for the generation
of functional expressions. The output is an expression for the
function which best fits the train set data.

In our case, the train set is the SNIa dataset, containing 307
pairs of the distance modulus $\mu(z)$ and the redshift $z$ as
described in Section 2. We also modify GDF to read an additional
file containing the errors of the corresponding data points, which
are used to evaluate the fitness function. No test dataset is
needed for our purposes.

The grammar used may involve standard numerical operations
($+,-,*,/$, etc.), as well as typical functions such as
$sin,cos,log,exp$ etc. and more complex production rules to turn
non-terminal expressions into terminal symbols. The choice of the
grammar vocabulary is apparently crucial for the behavior of the
algorithm, the speed of convergence and the quality of the final
fit. Our goal is to obtain a smooth fit for the distance modulus
curve in the range of $0\le z \le 1.75$, which will not lead to
problematic behavior upon differentiation and can produce a high
goodness of fit. The use of the $abs$ function is initially
excluded, since as it turns out is prone to lead to
discontinuities and unphysical behavior of the resulting dark
energy profile. Periodic functions may also lead to high frequency
oscillatory behavior in the candidate solutions, as the fitting
function attempts to interpolate between densely spaced
fluctuating datapoints. Solutions of this sort generally seem to
yield a better goodness of fit compared to grammars not employing
periodic functions. Such oscillatory behavior is hard to
distinguish within the accuracy available and there is little
theoretical justification to take it into account. We also find
that this family of grammars in the end yields higher errors for
the final model parameters, once a particular dark energy model is
tested using the periodic solutions. We thus choose to exclude
functions such as $sin$ and $cos$ for our final treatment of SNIa
data.

It is important to note that the method we discuss is highly
non-parametric. By this we mean that no parameters are used in the
process of fitting the initial dataset to obtain a candidate
solution. The GA does not apply some kind of minimization using a
seed parametric function. Instead, the maximization of the best
population fitness, which is equivalent to a $\chi^{2}$
minimization procedure, is accomplished in a purely stochastic
way. The resulting solution is given as an ordinary formal
mathematical expression of a function of $z$, in which no
parameters appear. This should be contrasted to other
non-parametric dark energy reconstruction methods, where some sort
of parameter minimization takes place from the very beginning,
even if these parameters are not directly related to some specific
dark energy model and only later do they get mapped to some
corresponding set of model-dependent quantities\cite{Daly:2003iy}. This is an
advantage of the GA approach, as it minimizes the amount of
initial model-dependent bias. Strictly speaking, the only
parameter which can be tuned beforehand is the vocabulary of the
grammar used. Grammatical evolution ameliorates this potential
source of bias, since it possesses  a far greater expressive power
than ordinary fit methods used, such as bin-wise polynomial fits.
The apparent drawback of the GA is that it is in the end overly
non-parametric, to the point where no established tools used in
usual minimization techniques can be employed to estimate and
propagate errors. We will discuss later how this problem can be
bypassed.
\begin{figure}[t!]
     \vspace{-1.15cm}\includegraphics[width=0.48\textwidth]{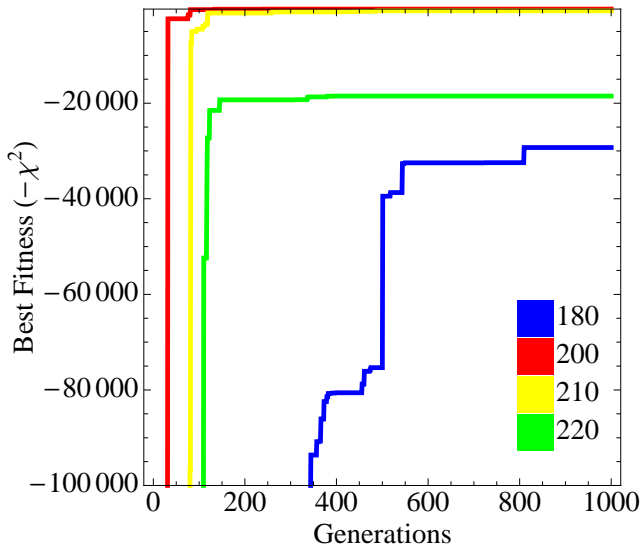}
    \label{fig2}
\caption{Convergence of the genetic algorithm as a function of the
number of generations. Curves are presented for different values
of the total number of chromosomes in the population. The case
with 190 chromosomes converges very slowly for the original SNIa
dataset and is omitted.}
 \end{figure}
The fitness function for the GA is chosen to be equal to
$-\chi^{2}_{SNIa}$, where \be \chi^2_{SNIa} = \sum_{i=1}^N
\frac{(\bar \mu_{obs}(z_i) -\bar \mu_{GA}(z_i))^2}{\sigma_{\mu \;
i}^2 }\,, \label{chi2ga} \ee and $\bar \mu(z)  = \mu(z)  - \mu
_0$. Notice the differences compared with (\ref{chi2}). The $
\chi^2_{SNIa}$ doesn't depend on parameters and $\bar \mu_{GA}(z)$
is the {\it reduced distance modulus} obtained for each chromosome
of the population through GE. We opt to fit for $\bar \mu(z)$
instead of $\mu(z)$, since the non-parametric approach we use
prevents us from marginalizing over $\mu_{0}$.

Another advantage is the fact that the GA does not require an
assumption of flatness as it can be seen from eq. (\ref{chi2}),
since it was used in order to find the reduced distance modulus
$\mu(z)$. Flatness (or non-flatness) comes into play when one
tries to find the luminosity distance $D_L(z)$ and the underlying
dark energy model given $\mu_{GA}(z)$. So, in this aspect the GA
is independent of the assumption of flatness as well.

The GA evaluates (\ref{chi2ga}) in each evolutionary step for
every chromosome of the population. The one with the best fitness
will consequently have the smallest $\chi^2_{SNIa}$ and will be
the best candidate solution in its generation. Of all the steps in
the execution of the algorithm, the evaluation of the fitness is
the most expensive. GDF is appropriately modified to use
(\ref{chi2ga}) as the basis of its fitness calculation.
\begin{figure}[t!]
     \includegraphics[width=0.5\textwidth]{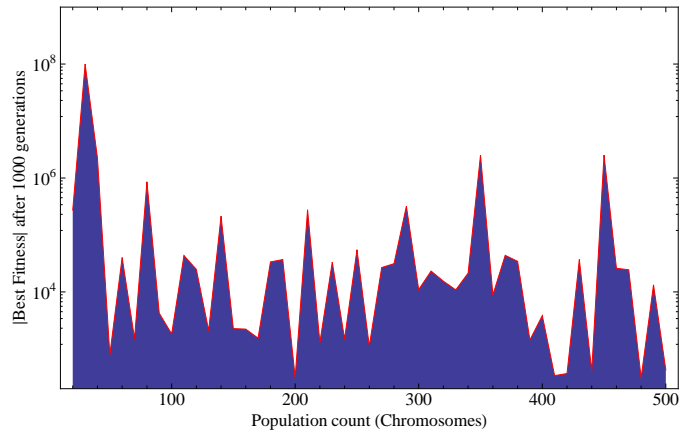}
    \label{fig1}
\caption{$|Best fitness|$ (lower is better) after $1000$
generations as a function of the chromosome count in the
population.}
 \end{figure}
Additionally, GDF requires a number of parameters which determine
the execution of the GA. These include the selection (percentage
of the population that goes unchanged into the next generation)
and mutation rates, the genome length of chromosomes in the
population, and the total chromosome count in one generation. A
maximum number of generations is also specified. We keep the
default values for the selection rate ($10 \%$) and mutation rate
($5 \%$).

 \begin{figure*}[t!]
     \includegraphics[width=0.48\textwidth]{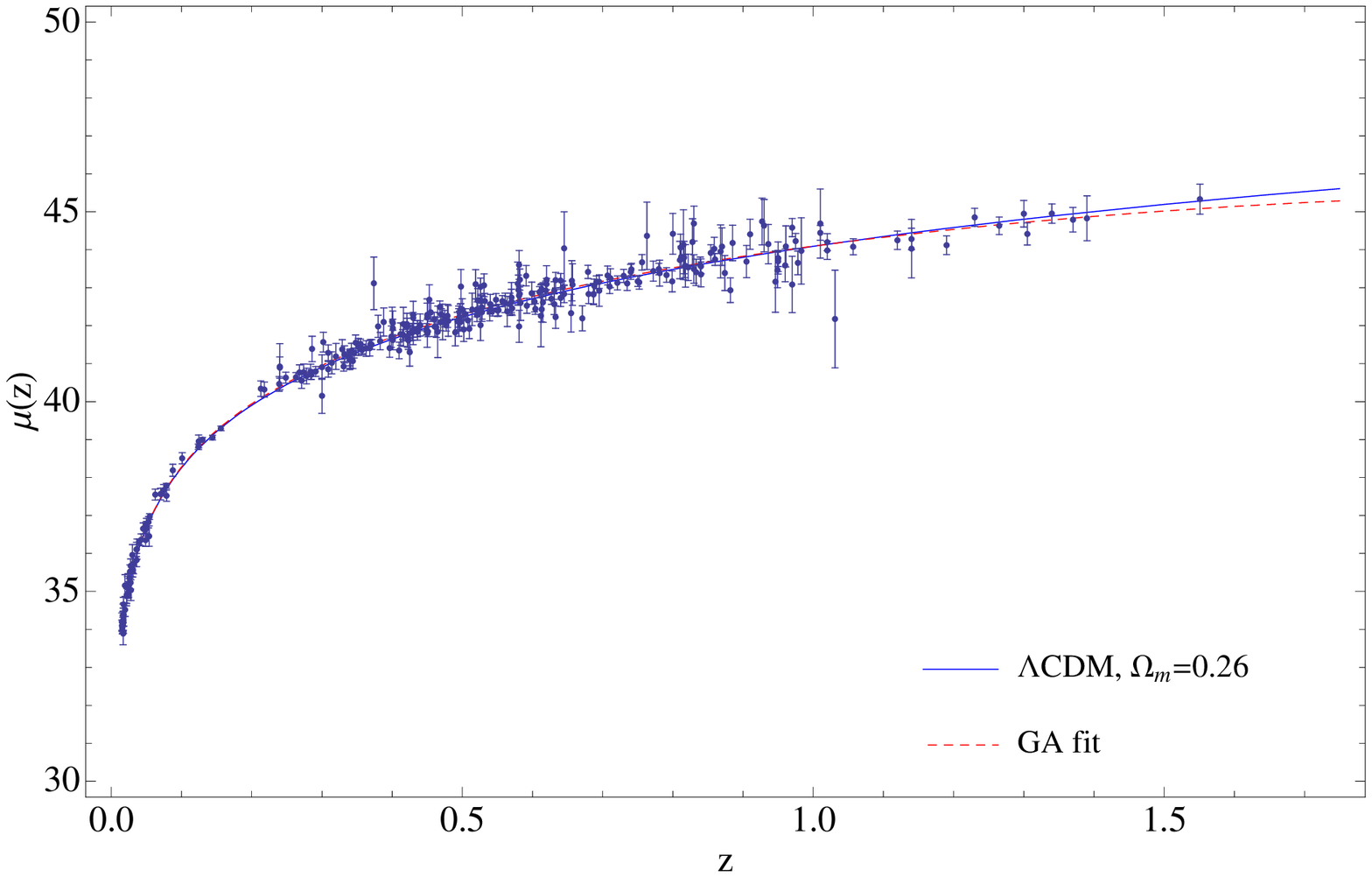}
     \includegraphics[width=0.48\textwidth]{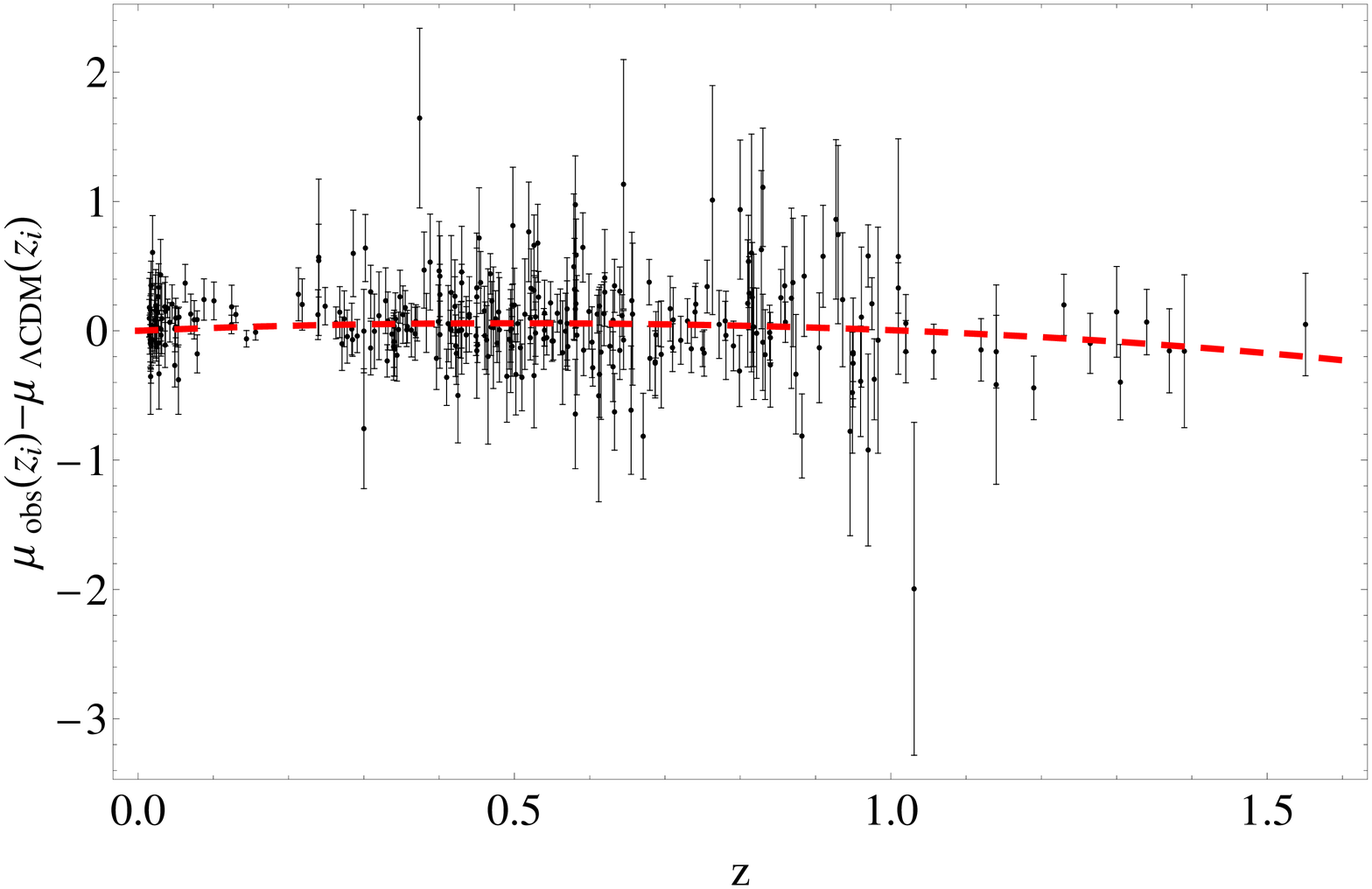}
    \label{fig2}
\caption{Left: The SNIa distance modulus against the redshift. The
solid line represents the $\Lambda CDM$ curve with
$\Omega_{m}=0.26$. The dashed line is a GA best fit function of
eq. (\ref{mzga}). Right: The data and the GA best-fit of eq.
(\ref{mzga}) residuals relative to $\Lambda CDM$.}
 \end{figure*}

The parameters which can significantly change the speed of the
algorithm is the genome length and the population size. Given the
use of GE, one should bear in mind that very long chromosomes with
many genes are not always fully used in the transcription process
to produce a formal expression. As such, we deem a chromosome
length of $100$ genes long enough, leading to solutions with a
high goodness of fit, without being computationally very
expensive. Having fixed the chromosome length, we must choose an
appropriate population size. Since the fitness function has to be
evaluated for every chromosome in one generation, this is the most
important parameter for the speed of execution. The GA is a
stochastic algorithm and thus the speed of convergence to an
optimal solution cannot be easily determined and is affected by a
number of parameters.

To find the most suitable population size, we make a preliminary
scan of this parameter space for different number of chromosomes
and determine what is the best fitness after $1000$ generations.
We determine sweet spots for $200$, $410$, $420$, $440$ and $480$
chromosomes. The first of them is used as the chromosome count for
the rest of the treatment, as it gives the fastest execution with
good results. All tests use an evolution of $1000$ generations.

After the execution of the GA, we obtain an expression $\bar
\mu_{GA}(z)$ for the reduced distance modulus as the solution of
best fitness and the corresponding  $\chi^2_{SNIa}$. Using this we
can, through differentiation, obtain parameters of interest, such
as dark energy equation of state parameters in the context of some
model. As already mentioned, this method gives no direct way to
estimate the errors for the derived parameters. One cannot expect
to get an estimate by just running the algorithm many times and
obtaining slightly different parameters. The GA usually tends to
converge at the same solution for a given dataset, unless we
change significantly the population size or the number of
generations. A way to circumvent this problem is to use Monte
Carlo simulation to produce synthetic datasets and rerun the
algorithm on them. We can thus obtain a statistical sample of
parameter values which will allow us to estimate the error. We
sketch the procedure we follow:
\begin{enumerate}
\item The GA is applied on the original SNIa dataset with the
chosen execution parameters and a solution for $\bar \mu_{GA}(z)$
is obtained. \item Choosing a particular dark energy model, we
determine its DE parameters from $\bar \mu_{GA}(z)$ using
differentiations. These parameters will be used as seed values to
produce synthetic datasets. \item A large number of synthetic
datasets is produced from the DE model, using the seed values from
the previous step. We add gaussian noise to the data points to
simulate real observational data. The standard deviation for noise
is taken to be equal to that of the corresponding observational
point. \item The GA is rerun for the synthetic datasets. \item A
new set of parameter values is determined from the solutions of
the synthetic datasets. \item Outliers are eliminated using
recursive trimming to reduce variance of the statistical sample.
\item The final mean values and standard deviations of the DE
parameters are determined.
\end{enumerate}
Using the above steps, we can obtain error estimates for the DE
model we wish to examine. Step 6 is essential to reduce the number
of outliers in the sample. The GA is stochastic and non-linear, so
adding gaussian noise to the input data doesn't lead to a gaussian
distribution of the output. For some of the synthetic datasets the
algorithm will fail to converge fast enough, yielding a suboptimal
solution, hence the presence of outliers. By recursively trimming
the sample, we can significantly improve the accuracy of the
result.

\subsection{Applications}
We will now give a number of examples where the method is applied
and the results interpreted in the context of some dark energy
model. We first use the GA to treat the more popular $\Lambda CDM$
model, which is also an easier example to demonstrate the method.
In the next section we consider a model with a dark energy
equation of state parameter which depends on the redshift and
determine the appropriate DE parameters of the model.

\subsubsection{$\Lambda CDM$}

For $\Lambda CDM$, the properties of dark energy are fully
determined, since $w=-1$ and the only free parameter of the model
is the total matter density $\Omega_{m}$. As described above, we
first run the algorithm on the SNIa dataset to obtain a seed
solution for the reduced distance modulus, and thus the distance
modulus $\mu_{GA}(z)$. Applying the GA to the original dataset as
it is was found to produce solutions with unphysical behavior for
small redshift, where we would expect $H(z) \to H_{0}$. For this
reason, a fiducial point was added in the dataset at small $z$
with a very low standard deviation, so that every good candidate
solution should take it into account. The total $\chi^{2}$ during
the GA runs included this artificial point. However, in our
subsequent treatment and final results we reverted to the initial
dataset (all $\chi^{2}$ values mentioned later do not include the
fiducial point). Using this method, the distance modulus was found
to be \be \mu_{GA}(z)=\log \left(z^{2.17145}
\left(-z^{2.82}+z+e^z\right)\right)+\mu_0\,. \label{mzga}\ee Note
that eq. (\ref{mzga}) is not unique since each time the GA is ran
it may give a different but equally good and model independent
approximation to the reduced distance modulus $\mu(z)$. This
function is then fitted against the corresponding modulus
predicted by $\Lambda CDM$ \be \mu _{\Lambda CDM} \left( z \right)
= 5\log _{10} D_L \left( {z;\Omega _m } \right) + \mu _0\,,
\label{LCDMmu}\ee with \be H \left( z \right)^2 =
H_0^2\left[\Omega _m \left( {1 + z} \right)^3  + 1 - \Omega
_m\right] \,, \ee to obtain the seed value of $\Omega_{m}$. We
then use the Monte Carlo method to produce $100$ synthetic
datasets from (\ref{LCDMmu}) adding gaussian noise and rerun the
GA on them. The resulting set of $\mu_{GA}(z)$ from the synthetic
datasets is used to obtain in the same fashion a statistical
sample of $\Omega_{m}$ values. Due to the non-linearity of the GA
and GE implementation, the sample is not normally distributed and
contains a lot of outliers. This is easy to establish both
graphically through histograms, or by comparing the mean and
standard deviation of the sample with its corresponding robust
estimators, the median and median deviation. The large discrepancy
between these quantities is a sign of outliers in the dataset. We
eliminate these by recursively removing outlier points. The
heuristic we use is to discard in each step data points that are
more than 2$\sigma$ away from the mean of the distribution at that
stage. The trimming continues until we reach a stationary state,
where no other points are removed. From the improved statistical
sample we determine the final values of the mean and standard
deviation for the dark matter density to be $\Omega_{m}=0.308 \pm
0.029$, in good agreement with the WMAP value of $\Omega_{m}=0.258
\pm 0.030$ \cite{Komatsu:2008hk}.

\vspace{0pt}
\begin{table}[!b]
\begin{center}
\caption{Comparison of the mean and standard deviation of the
$\Omega_{m}$ statistical sample produced from the synthetic
datasets and the corresponding median and median deviation. The
initial large discrepancy is greatly reduced by trimming outlier
points.}
\begin{tabular}{ccccccc}
\hline
\hline\\
\vspace{6pt}   & Mean                      & Median                         & $\Delta$ ($\%$)    & $\sigma$ &$\sigma_{median}$ & $\Delta$ ($\%$)     \\
\hline \\
\vspace{6pt}  \textbf{Initial}& $0.633$ & $0.357$& 77.5  & $0.584$ & $0.074$ & 692 \\
\vspace{6pt}  \textbf{Trimmed} & $0.308$ & $0.298$&  3.26 & $0.029$ & $0.016$ & 83.3 \\
\hline \hline
\end{tabular}
\end{center}
\end{table}

\begin{figure}[t!]
\centering
   \begin{minipage}[c]{0.25\textwidth}
   \centering \includegraphics[width=1\textwidth]{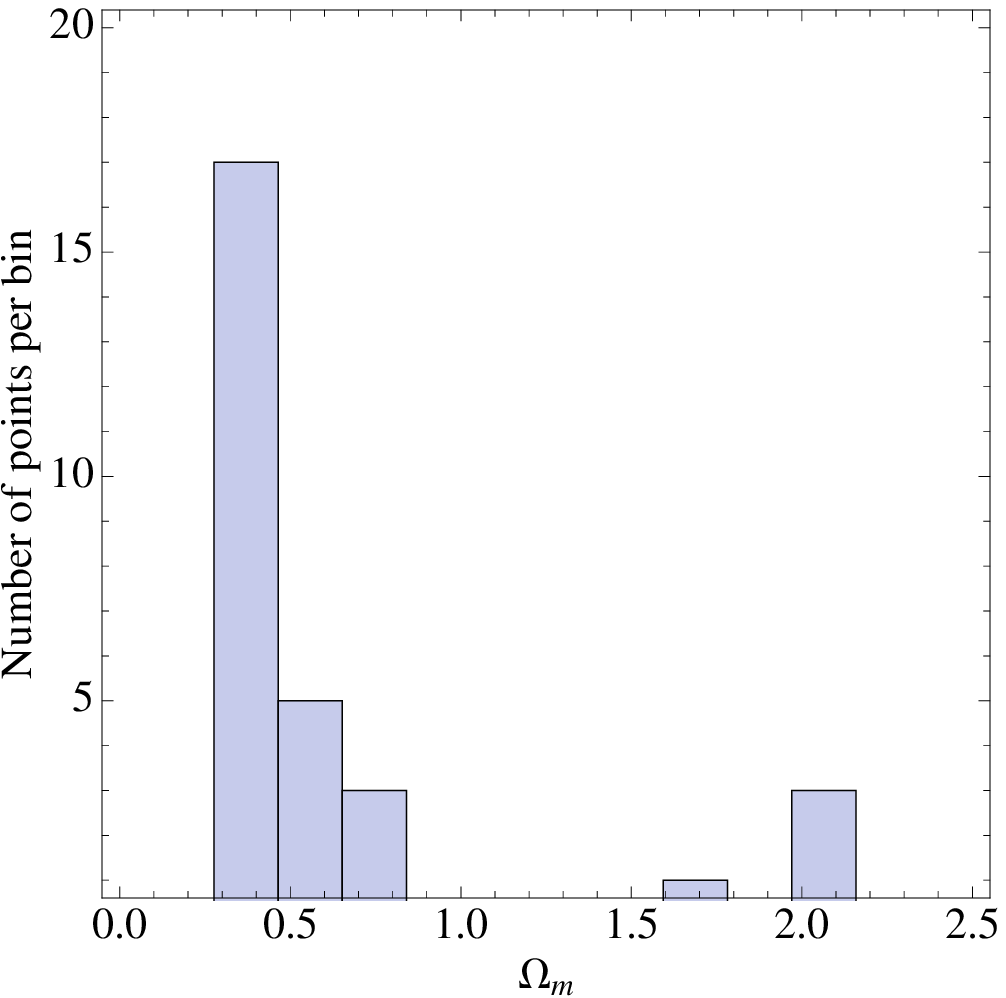}
   \centering \includegraphics[width=1\textwidth]{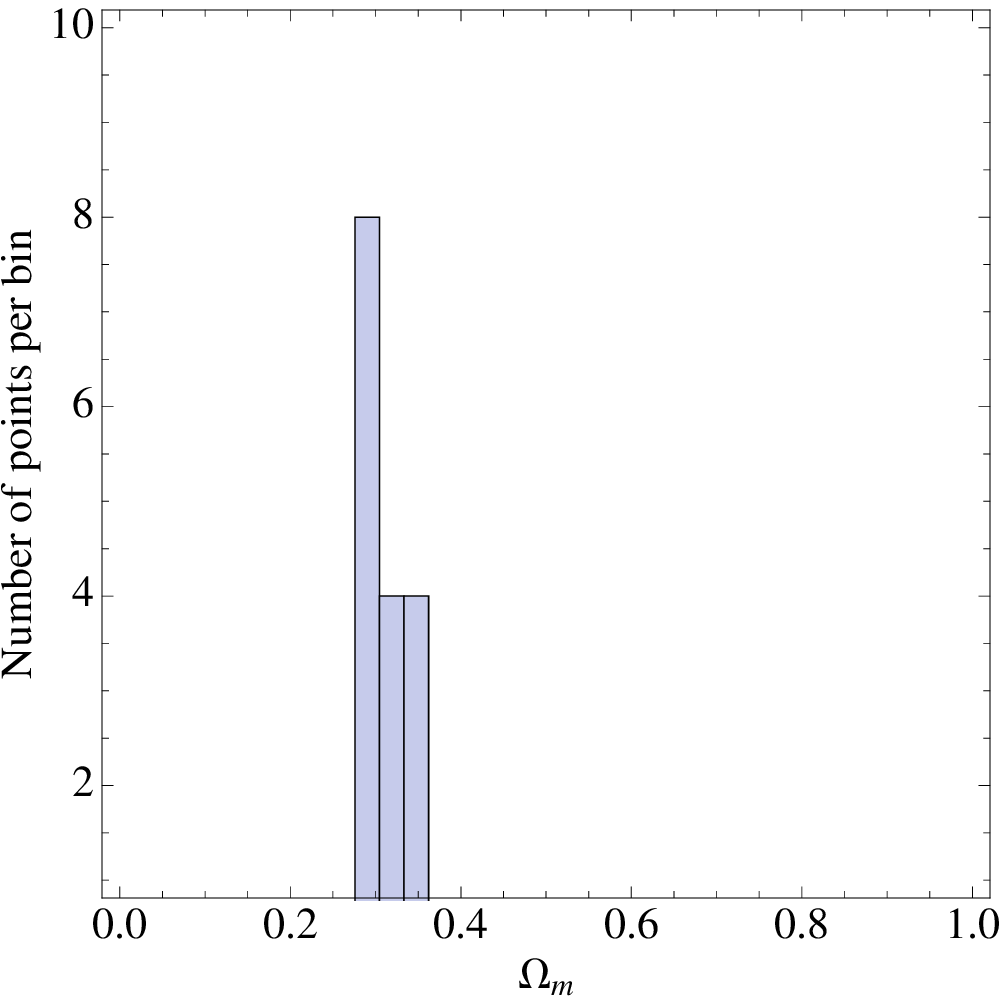}
   \end{minipage}
   \caption{Histograms showing the dispersion of the set of $\Omega_{m}$ values.
   Upper: Initial set. Lower: After the recursive trimming of outliers.}
   \label{figure2}
\end{figure}%

\begin{figure*}[t!]
\centering
   \begin{minipage}[c]{1\textwidth}
   \centering \includegraphics[width=0.32\textwidth]{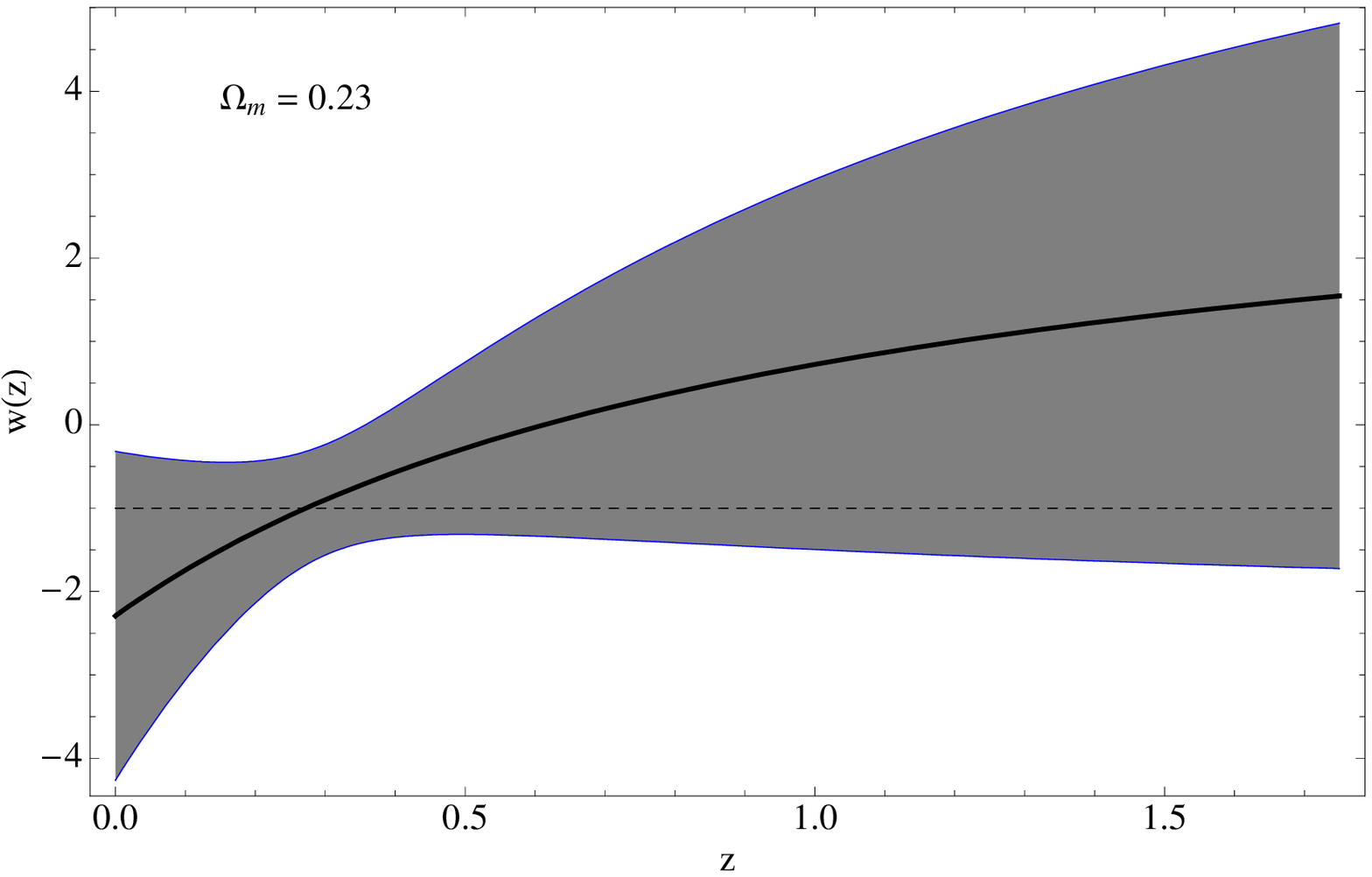}
   \centering \includegraphics[width=0.32\textwidth]{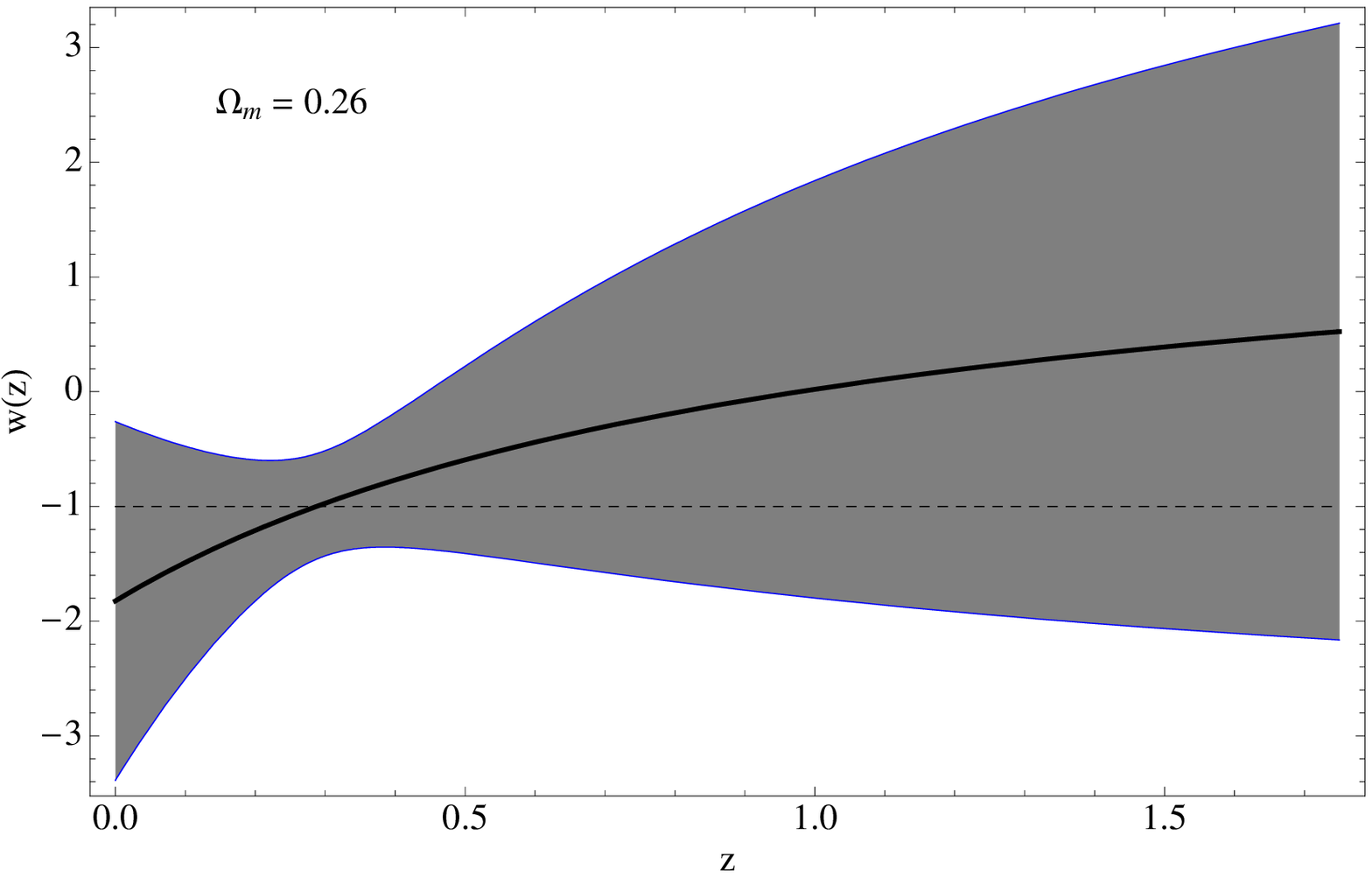}
   \centering \includegraphics[width=0.32\textwidth]{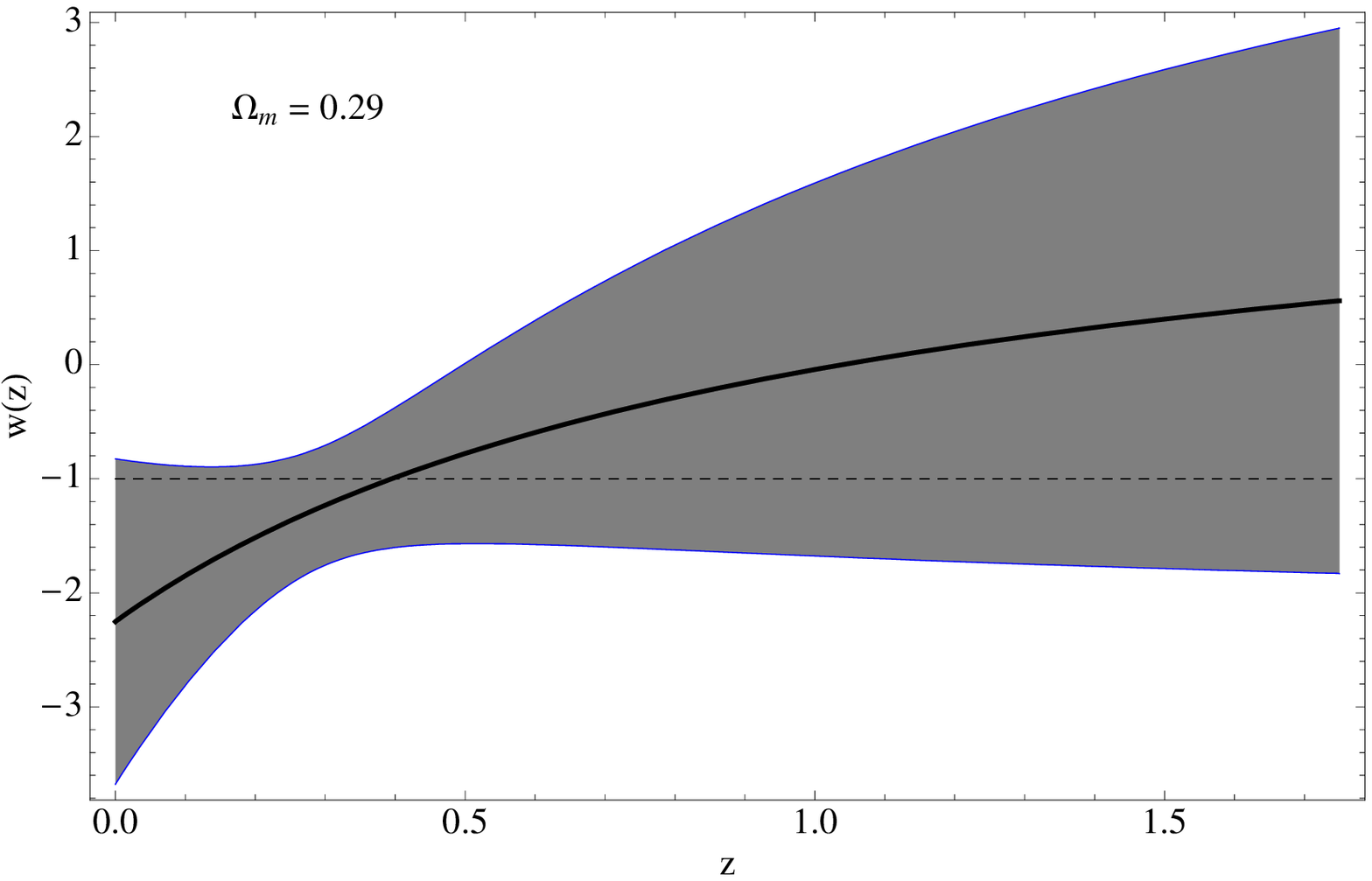}
   \end{minipage}
   \caption{The DE equation of state $w(z)$ for different values of $\Omega_{m}$.}
   \label{figure4}
\end{figure*}%

\vspace{0pt}
\begin{table*}[!t]
\begin{center}
\caption{Values for mean and median and their corresponding errors
for the Chevallier-Polarski-Linder DE ans\"atz. Large
discrepancies between the mean and the median signal the presence
of outliers. Overall $\chi^{2}$ can be significantly -and
adversely- affected by such points. The trimmed samples generally
present a considerably better behavior.}
\begin{tabular}{ccccccccccc}
\hline
\hline\\
\vspace{6pt}  & $\Omega_{m}$ & Mean $w_{0}$  & Median $w_{0}$   & $\Delta w_{0}$ ($\%$)    & $\Delta \sigma w_{0}$ ($\%$) & Mean $w_{1}$  & Median $w_{1}$       & $\Delta w_{1}$ ($\%$)    & $\Delta \sigma w_{1}$ ($\%$) & $\chi^{2}$    \\
\hline \\
\vspace{6pt}  \textbf{Initial}& 0.23 &$-1.8 \pm 6.9$ & $-2.4 \pm 1.3$& 24.6  & $424$ & $4 \pm 25$ & $5.5 \pm 5.5$ & $23.5$& $345$ & 352.2 \\
\vspace{6pt}  \textbf{Trimmed}& 0.23 &$-2.3 \pm 2.0$ & $-2.4 \pm 1.1$& $4.7$  & $73$ & $6.0 \pm 7.9$ & $5.6 \pm 3.8$ & $7.7$& $109$ & 471.0 \\
\vspace{6pt}  \textbf{Initial}& 0.26 &$-9 \pm 69$ & $2.4 \pm 1.5$& 278  & $4480$ & $30 \pm 240$ & $5.3 \pm 7.5$ & $456$& $3140$ & 3797 \\
\vspace{6pt}  \textbf{Trimmed}& 0.26 &$-1.8 \pm 1.6$ & $-2.2 \pm 1.1$& $18.4$  & $37.8$ & $3.7 \pm 6.5$ & $5.2 \pm 4.7$ & $28.8$& $37.8$ & 343.4 \\

\vspace{6pt}  \textbf{Initial}& 0.29 &$-50 \pm 470$ & $-2.4 \pm 1.2$& $1960$  & $38500$ & $200 \pm 1700$ & $4.6 \pm 6.6$ & $3810$& $25500$ & 6667 \\

\vspace{6pt}  \textbf{Trimmed}& 0.29 &$-2.3 \pm 1.4$ & $-2.5 \pm 0.8$& $7.9$  & $75.5$ & $4.4 \pm 5.5$ & $5.0 \pm 3.9$ & $12.1$& $48.8$ & 420.6 \\
\hline \hline
\end{tabular}
\end{center}
\end{table*}

\subsubsection{Chevallier-Polarski-Linder DE ans\"atz}
In this dark energy model, the equation of state parameter is
given by the ans\"atz,\cite{Chevallier:2000qy,Linder:2002dt} \be
w\left( z \right) = w_0  + w_1 \frac{z}{{1 + z}}\ee with two
parameters $w_{0}$ and $w_{1}$ determining the temporal profile of
$w(z)$. This model has been extensively used in previous
parametric reconstructions of dark energy from SNIa data. The
distance modulus of the model is given by equations (\ref{mdl})
and (\ref{dlth1}) using \bea H \left( z
\right)^2 &=& H_0^2[\Omega _m \left( {1 + z} \right)^3  \\
&+& \left( {1 - \Omega _m } \right)\left( {1 + z} \right)^{3\left(
{1 + w_0 + w_1 } \right)} e^{ - 3w_1 \frac{z}{{1 + z}}}] \,.\eea
Here we use the matter density as an input parameter and attempt
to reconstruct the dark energy profile by determining the two DE
parameters $w_{0}$ and $w_{1}$. We repeat the process three times
using as $\Omega_{m}$ the WMAP value $\pm 1\sigma$. As we see in
Table 2, the WMAP value gives the best goodness of fit. The method
is roughly the same as for $\Lambda CDM$, only now we have to
determine two parameters. After running the GA on the SNIa dataset
and getting the seed values for $w_{0}$ and $w_{1}$, we produce
$100$ synthetic datasets and reapply the algorithm to get a sample
of parameter values. Again, we have to remove outlier points to
decrease the dispersion of the data. To do so, we recursively
remove data points lying outside the $90\%$ confidence level in
every step, until we reach a stationary state. From the improved
sample values we get the mean and errors for $w_{0}$ and $w_{1}$.

The profiles we get for the DE equation of state parameter are in
accordance with previous parametric reconstructions based on this
model. We see in Fig. 5 there is a trend for a redshift-varying
$w(z)$, with a crossing of the $w=-1$ line at late times. Still,
the errors are large for the method to make any conclusive
arguments about these effects. The errors are essentially due to
the limited number of MC synthetic datasets used for this
treatment. Increasing the number of datasets would lead to better
statistics and decrease the errors further, but at a great
computational cost, since the GA should then be applied to a much
larger collection of data.

\section{Conclusions}

We presented a method for non-parametric reconstruction of the
dark energy equation of state parameter based on the genetic
algorithms paradigm and grammatical evolution. The method was
applied to determine parameters of cosmological models such as the
$\Lambda CDM$ and the Chevallier-Polarski-Linder DE ans\"atz. This
approach produces results that are in agreement with previous
parametric and non-parametric reconstructions. The main advantage
of the GA is the absence of any parametrization in the treatment
of the initial data which can reduce any model-dependent bias. In
our case the desired function to be found is the reduced distance
modulus $\mu(z)$. It should be noted that running the GA several
times may give different but equally good model independent
approximations to the distance modulus $\mu(z)$, so what we
actually need some statistics and a model to interpret the
results. The reason for this is that if we want to constrain a
parameter like $Omega_m$, then already we have made a (minimal)
assumption for the underlying theory. A similar approach has been
also used in the literature before, see for example
\cite{Allanach:2004my} where the authors use the GA to
discriminate between the various SUSY models.

On the other hand, the shortcomings of the method include
computational speed issues, since the execution of the GA is
usually very expensive regarding the CPU time needed. This in turn
limits the amount of precision one can obtain using Monte Carlo
methods to estimate errors for the final model parameters.
However, this issue can soon be resolved with the advent of faster
systems and more efficient GA implementations. Other open issues
include the determination of optimal algorithm parameters for the
SNIa dataset, as well an extension of the method to analyze a
larger variety of dark energy models.

\section*{Acknowledgements}
We would like to thank I. Tsoulos for helpful discussions and
suggestions on genetic algorithms and grammatical evolution
programming. The tool GDF is available from
\href{http://cpc.cs.qub.ac.uk/summaries/ADXC}{http://cpc.cs.qub.ac.uk/summaries/ADXC}.
C.B. is supported by the CNRS and the Universit\'e de Paris-Sud
XI. S.N. acknowledges support by the Niels Bohr International
Academy, by the EU FP6 Marie Curie Research $\&$ Training Network
``UniverseNet" (MRTN-CT-2006-035863) and from the Danish Research
Council (FNU grant 272-08-0285).

\end{document}